An Investigation of Three-point Shooting through an Analysis of NBA Player Tracking Data

By

Bradley A. Sliz

Thesis Project

Submitted in partial fulfillment of the

Requirements for the degree of

MASTER OF SCIENCE IN PREDICTIVE ANALYTICS

December, 2016

Dr. Alianna JeanAnn Maren, First Reader

Thomas Robinson, Second Reader

## Abstract

In my thesis, I address the difficult challenge of measuring the relative influence of competing basketball game strategies, and I apply my analysis to plays resulting in three-point shots. I use a glut of SportVU player tracking data from over 600 NBA games to derive custom position-based features that capture tangible game strategies from game-play data, such as teamwork, player matchups, and on-ball defender distances. Then, I demonstrate statistical methods for measuring the relative importance of any given basketball strategy. In doing so, I highlight the high importance of teamwork based strategies in affecting three-point shot success. By coupling SportVU data with an advanced variable importance algorithm I am able to extract meaningful results that would have been impossible to achieve even 3 years ago.

Further, I demonstrate how player-tracking based features can be used to measure the three-point shooting propensity of players, and I show how this measurement can identify effective shooters that are either highly-utilized or under-utilized. Altogether, my findings provide a substantial body of work for influencing basketball strategy, and for measuring the effectiveness of basketball players.

# Acknowledgements

Firstly, I would like to express my sincere appreciation to my thesis committee, Dr. Alianna Maren and Thomas Robinson.  Their patience and support was a beacon that helped guide me to the end.  Thank you!

Lastly I would like to express my deepest gratitude to Dr. Rajiv Shah.  His advice, collaboration, and excitement provided me the energy needed to finish my thesis amid the stresses of family life and full time employment.  Without his friendship and cooperation, my work would have been only a shell of what it is.  Thank you!

# Table of Contents

# Introduction

Basketball is a game of athleticism, skill, positioning, and teamwork. Teams that optimize each of these facets of their game can generally expect to be successful. However, it is difficult to measure the degree to which a given strategy can influence basketball success, because there are many competing influencers (i.e. did a player make a shot because they were open, or because they are a good shooter?), and because there is so much noise mixed in with the signal (i.e. even great three-point shooters only make 40% of their shots).

With the advent of player tracking data, it has become possible to explore game strategies in a new light. Player tracking data enables measurements that were not before measureable beyond subjective suppositions and terse remarks. In fact, across sports, player tracking is revolutionizing the sports-analytics movement with copious collections of fine-grained game observations, enabling an assortment of (literally) game-changing analyses. In basketball research, much work has been done to leverage player tracking data, but little work has used it to analyze three-point shooting.

In my thesis:

- I analyze player tracking data from over 600 games from the first half of the 2015-2016 NBA season, to find plays resulting in three-point shots.
- I derive custom position-based features that capture tangible game strategies from game-play data.

- I propose statistical methods for measuring the relative importance of any given basketball strategy.

- I demonstrate how these position-based features can be used to measure the three-point shooting propensity of players.

- Finally, I show how this propensity metric can identify effective shooters that are either highly-utilized or under-utilized.

# Background

Between 2010 and 2013, the NBA equipped all of its arenas with motion capture cameras. Throughout the subsequent basketball seasons, positional data were collected in every regular season and post season game. During each game, the positions of the ball and each player on the court were recorded at a rate of 25 observations per second. This rich dataset has enabled researchers, analysts, and basketball aficionados alike to explore the game of basketball in ways that were never before possible.

YonggangNiu [2014] offers an excellent description on the background of the technology that enables the collection of this data in their paper *Application of the SportVU Motion Capture System in the Technical Statistics and Analysis in Basketball Games.* The following paraphrases the discussion in that paper on the SportVU technology:

> The SportVU system (Multi-lens Tracing System) was invented in 2005, by Israeli scientist Mickey Tamir, and was originally intended for missile tracking in a military setting. The technology was also shown to have functional applications in sports. In 2008, the sports analytics firm STATS acquired the SportVU technology and focused it on the analysis of basketball games. Today, this system has been installed in every NBA teams' home court and has captured motion data for over 1000 professional basketball games.

> To date, this NBA SportVU data has already occupied an important position in the academic world. The annual Sloan Sports Analytics Conference at the Massachusetts

Institute of Technology is the top technology event in the sports world. Among the papers submitted to Sloan about basketball last year, half were based on the data captured for the NBA by the SportVU system.

The SportVU system is run by STATS Data Corporation Limited. The ceiling of every basketball gymnasium in the NBA is equipped with 6 cameras and every half-court has 3 cameras, all synchronized to each other. Collectively, these cameras capture player and ball movements, and extract XYZ locations relative to the court at a rate of 25 frames per second. Furthermore, these positional data are collected with a foreign key that can be used to join onto each game's Play-by-Play records.

This decision by the NBA to equip all of its arenas with STATS SportVU systems was pivotal in ushering in a new age of data driven strategy to the game of professional basketball.

# Review of the Literature

## SportVU

In his paper *CourtVision: New Visual and Spatial Analytics for the NBA*, Goldsberry [2012] proposed the use of spatial analytical techniques to assess NBA player's shooting abilities. His work was one of a number of efforts beginning to challenge box-score analytics as the status quo for basketball performance assessment. He suggested that spatial analysis was vital to the study of NBA basketball, and this suggestion has only become more true in the past five years. Indeed, his work helped pave the way for the NBA to buy in to collecting player tracking data with STATS SportVU, which spawned a flurry of in-depth NBA spatial analyses that continue to contribute substantially to the domain of basketball analytics.

With the advent of STATS SportVU tracking data in the NBA, basketball researchers have been able to explore in-game interactions, strategies, and player performance in innovative ways that have not before been possible. Specifically, the granularity at which the SportVU data are collected enable a precision of measurement that before was not possible in analyzing the game of basketball. Indeed, in the four years since Goldsberry's seminal work, the field of basketball analytics has been revolutionized by analytics with SportVU tracking data. It has been leveraged to inform all facets of the game, from team member selection, to team strategy, to player development. The following are some examples of this radical re-envisioning:

- Cervone et al. [2014] demonstrate that ***player-tracking data can be leveraged to***

***evaluate every decision made during a basketball game***, whether it be to pass, dribble, shoot, etc. Furthermore, they show that by applying their modeling framework to every moment (25 frames per second) of a basketball game, a multitude of new metrics and analyses of basketball become feasible; they offer some examples of these new metrics for answering real basketball decisions.

- In a more recent paper, Cervone et al. [2016] expand on their previous work to show how ***new positional-based metrics can be leveraged to influence basketball strategy***. They use SportVU tracking data to assess the value of the spatial regions of the basketball court. They infer the value of court real estate based on player and ball movement alone. As in their previous work, they develop new metrics for assessing both offenses and defenses at the player and team levels.

- Maheswaran et al. [2014] show that ***simple basketball statistics such as rebounds can be observed in much more complex ways than simply numbers in a box-score***. They use player tracking data to deconstruct rebounds into subcomponents that help to better explain rebound events. They propose that a rebound can be considered from three distinct dimensions: Positioning, Hustle and Conversion, and that player tracking data can enable rebound events to be observed in these contexts. Like Cervone, they demonstrate how sports tracking data can enable the creation of novel metrics for evaluating the game of basketball.

- Lucey et al. [2014] use ***player tracking data to explain how shooters get open***. First, they confirm the notion that on-ball defensive pressure reduces shooting percentages. Given this, they investigate how an offense can get shooters open. They demonstrate that the frequency of defensive role-swaps is predictive of open shots, and use this finding to measure teams' defensive effectiveness. Furthermore, they describe a method that can be used to query similar historical plays by using tracking data as the query input.

Remarkably, this is only a small sample of the work done to date that has demonstrated the value of SportVU data. More recent research is pushing its limits even farther, from automatic play categorization, to applications with neural networks, to the prediction of injuries before they happen. Truly, the uses of SportVU data are bountiful. More significantly, SportVU data is enabling sports analyses that are both unique and meaningful to the game of basketball. Here are a few exceptional examples:

- McIntyre et al. [2016] propose that their work can be ***consumed as one component of a coaching assistance tool for analyzing plays***. They use player tracking data to train a classifier that labels ball screen plays according to common defensive response strategies: Over, Under, Trap, and Switch.

- Wang and Zemel [2016] demonstrate how ***long short term memory (LSTM) recurrent***

**neural networks can consume voluminous amounts of the fine-grained SportVU data** to perform analyses and comparisons of basketball plays that would not be possible for a human observer alone. They focus on the classification of offensive plays. The use of an LSTM allows their network to learn the complex interactions between all the players on the court as they evolve over the course of a play. Furthermore, they show how their model can still perform well when trained on one season and tested on the next.

- Talukder et al. [2016] present a model that uses **SportVU player tracking data to predict the likelihood that any given player will sustain an injury during the course of an upcoming game**. They combine play-by-play game data, SportVU data, player workload and measurements, and team schedules to train their predictive model. They argue that by combining their results with information on team schedules and rest days, teams can identify the best time to rest their star players and reduce long-term injury risk. This work is significant because it demonstrates how player tracking data can impact the game beyond just basketball strategy; it can be harnessed to manage player health, and, by association, fan interest and revenue. Furthermore, it can be used by fantasy sports fans to manage their own investment risks.

In sum, there is a substantial body of work developed in the last few years encompassing the analysis of basketball with NBA SportVU tracking data. Because positioning is so central to the game of basketball, Goldsberry's [2012] suggestion is becoming more and more true: spatial analysis is vital to the study of the game. The flood of data collected during games via SportVU is revolutionizing basketball analytics.  This revolution is challenging core principles of the game including game strategy, performance assessment, and team and player management.  Likewise it as an exciting time to be involved in basketball research because each new innovation opens doors to many new analyses and poses questions about how we understand the game.

**Variable Importance**

Important variable measurement is a key component of this work, so consider some background on this topic.  Some of the most commonly used machine learning algorithms such as random forests and gradient boosting machines provide measures for predictor variable importance along with their resultant models.  Breiman [2001] discusses variable importance in his *Random Forests* paper.  He describes how out-of-bag predictors are randomly permuted to measure percent increase in misclassification rate for each predictor variable, to give a strong estimate of variable importance for the given classification or regression task.  He also describes how random forests are robust to collinearity, and can implicitly capture variable interactions in their variable importance measurements.  Since their introduction, random forests have become a standard method for measuring important variables. Given their strengths, random forests may be a perfect vehicle for assessing basketball strategies in my work.

However, random forests do have some flaws in variable importance measurement.  In their paper *Bias in random forest variable importance measures: Illustrations, sources and a solution*, Strobl et al. [2007] discuss how random forests are not reliable in situations where predictor variables vary in their scale of measurement or their number of categories. Specifically, they demonstrate that when random forest variable importance measures are used with data of varying types, the results are misleading because suboptimal predictor variables may be artificially preferred. They propose conditional inference forests as a strategy to counteract these biases.

One downside to the conditional inference forests proposed by Strobl et al. [2007] is computational inefficiency, so I consider an alternative method for my work. In their paper *Feature Selection with the Boruta Package,* Kursa and Rudnicki [2010] describe how their algorithm Boruta controls for the variable importance biases of a random forest. Specifically, they standardize importance measures to z scores, and intentionally include features in the model that are random by design; these are known as 'shadow' features. A shadow feature's Boruta importance score can be nonzero only due to random fluctuations. Thus the set of importance scores of shadow features is used as a reference for deciding which actual features are truly important. Effectively, anything that performs worse than these shadow features is considered no better than random. Further, the Boruta algorithm implementation is efficient enough that dozens of iterations can be performed on my data to assemble feature importance distributions, rather than merely scalar measurements.

In my work, I use the Boruta algorithm to measure variable importance, because it is a more advanced (and more current) method which can overcome the deficiencies of random forest variable importance measurement for my problem. By coupling the most recent innovation in basketball data-gathering (SportVU), with an advanced variable importance algorithm (Boruta), I am able to extract meaningful results that would have been impossible to achieve even 3 years ago.

# Methods

## Make-Miss Model

As a whole, this research investigates three-point shot strategies in the NBA. To accomplish this, three-point strategy is investigated from two different frames of reference. First, three-pointers are studied at the play level, where in-game strategies and actions are compared for their power at influencing three-point shot success. Specifically, a model is trained to measure each variable's importance in influencing a make or miss. To visualize this make-miss model, consider how each basketball game is made up of many plays, and how each play is made up of the actions of 5 players from each team and the ball, as depicted in Figure 1.

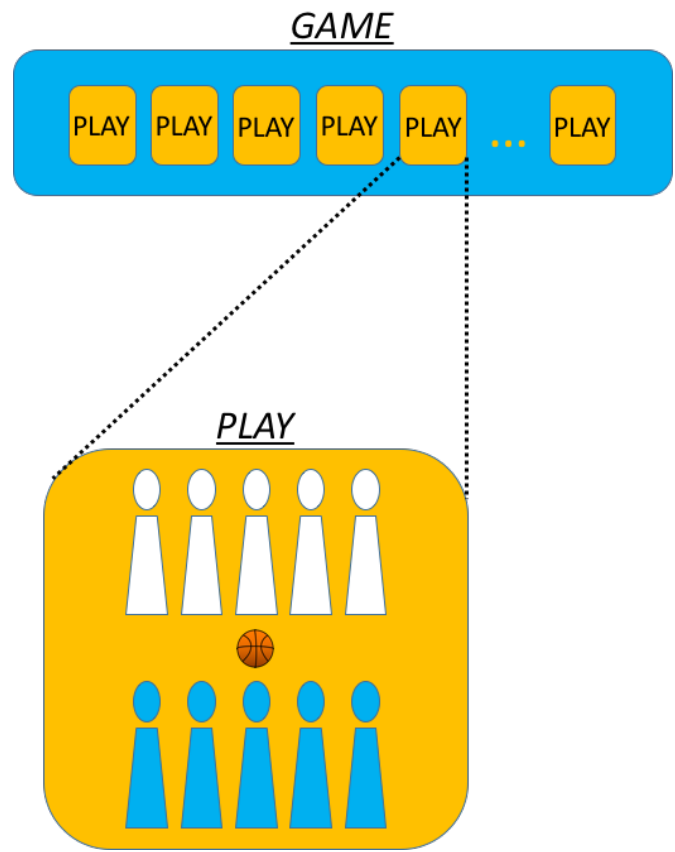

*Figure 1: Depiction of the make-miss model frame of reference*

As depicted in Figure 1, each play is made up of the actions of 5 players from each team and the ball. I use these player and ball actions to construct custom features that capture game strategy such as teamwork, player matchups, and on-ball defender distances. These features are aggregated into a single observation for each play, across all games. Likewise, the make-miss model is constructed on this collection of observations of my custom features for each play.

The structure of a basketball game lends itself perfectly to a classification problem, because every shot taken has a binary outcome: a make, or a miss. This analysis uses the play (specifically three-point plays) as its unit of measurement, and seeks to quantify the relative value of different offensive strategies at that play level. Likewise, the variable importance measures returned by the Boruta algorithm are perfect vehicles for quantifying the relative values of play strategies. By considering the make / miss of a three-pointer as a classification problem, I fit a model to predict the outcome of a play, then compare the importance of the dependent variables.

**Player Model**

To be competitive at making three-point shots in the NBA, understanding the relative strength of various game strategies and actions is a strong start. However, three-point shooting is a skill, and one that varies greatly even at the professional level. Likewise, it is highly valuable to assess three-point shooting across players.

The second frame of reference I use to analyze three-point shooting is at the player level, where the same in-game strategies and actions measured in the make-miss model are collapsed to comprehensive values for each player. To visualize the player model, consider how in each game, a given player may take a three-point shot on multiple plays. For each player, I aggregate all of their three-point shooting plays across all games, as depicted in Figure 2.

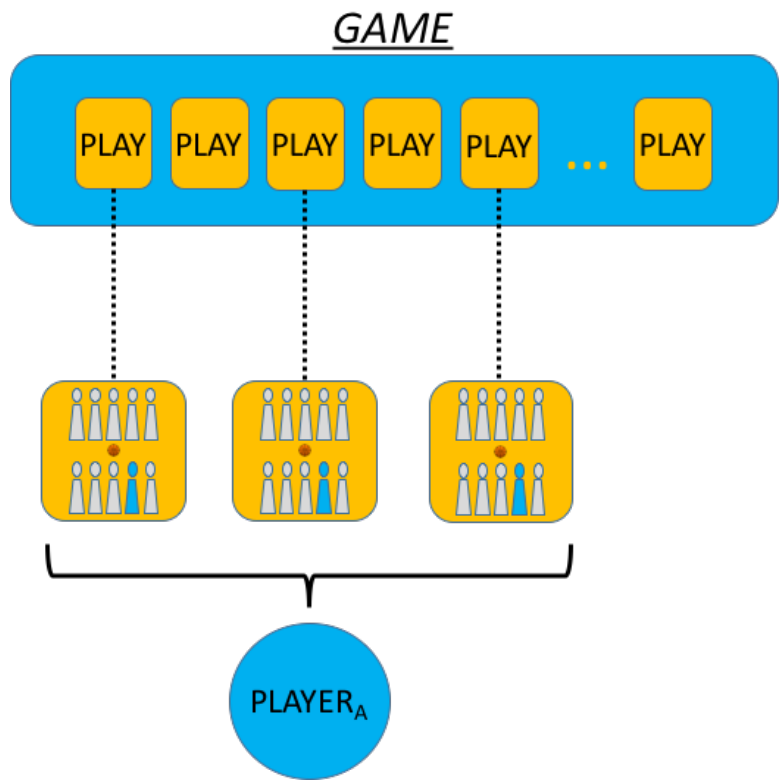

*Figure 2: Depiction of the player model frame of reference*

As depicted in Figure 2, player A shot three-pointers on multiple plays. I collect the make-miss model features for all three-point shooting plays for player A, across all games, and aggregate them to form a single observation for player A. I do this aggregation for all players who attempted a three-point shot. This collection of player observations forms the data on which I build the player model.

In the player model, I aggregate the metrics derived in the make-miss model to each shooter in my dataset to identify trends in player usage. By aggregating the features defined in the make-miss model, I am able to capture comprehensive measurements of the movement of players and their teams on their three-point shooting plays. Specifically, the player model uses a gradient boosting machine regression algorithm to predict three-point attempts. By comparing the model's prediction for a player's per-game three-point attempt rate to their actual three-point attempt rate, I can identify players who are behaving in unexpected ways. I quantify both the most effective shooters, and the most under-utilized shooters.

Next, consider the modeling strategy I deployed for the player model problem. A typical modeling framework might include a train dataset, and a test dataset, such that the train set is used to train the model, and the test set is used to evaluate the model's performance on unseen data. This architecture would look something like Figure 3, where the orange box represents training data containing observations for players $1$ through $n$, and the blue box represents testing data containing observations for players $m$ through $z$:

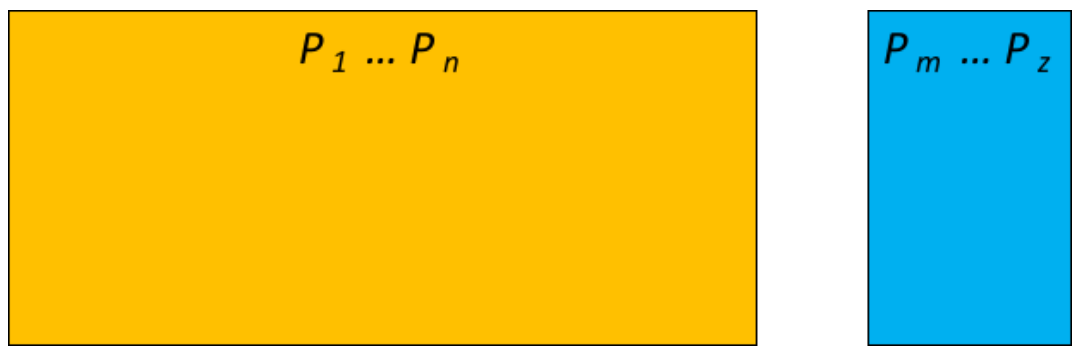

*Figure 3: Typical modeling data framework*

However, because each player in my dataset needs a prediction, this modeling methodology will not suffice. Instead, I deploy an iterative leave-one-out modeling approach on top of my train-test split. While the test set remains as an unseen holdout, the train set is split further, such that I train one model for each player in the train set, as in the following Figure:

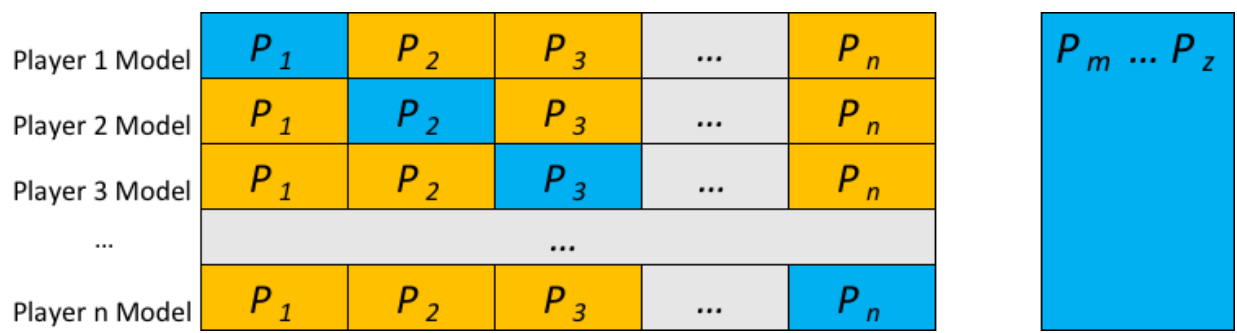

Model Training
Holdout

*Figure 4: Iterative leave-one-out modeling data framework*

The modeling architecture displayed in Figure 4 allows for every player to be scored on a model in which they were not included for training. This is important because it protects the player scores from being over-biased, as in a case where the model has already "seen" the player it is scoring. Also, by maintaining a holdout test set, I can evaluate the performance of every player's model and assess model consistency across the players; and because each player model's training set only differs by one observation, we can expect consistent model performance.

Next, consider the means by which players can be assessed based on the outputs of their respective player models. As I described above, I first find the deviation between the model's prediction for a player's per-game three-point attempt rate and their actual three-point attempt rate. In a sense, I use the error term of the regression model to identify players who are behaving in unexpected ways. Specifically, I measure player model deviation like this:

$$deviation = (actual\ 3PA) - (predicted\ 3PA)$$

In the above equation, deviation is defined as the difference between a player's actual three-point attempt rate, and their model-predicted three-point attempt rate. This deviation alone can identify players who shoot three-pointers more or less frequently than other players with like in-game experiences. However, as mentioned before, three-point success is highly dependent on player skill. Likewise, I propose a new metric for measuring a given player's three-point propensity, by applying a penalty on deviation according to the player's three-point shooting percentage. Specifically, I measure propensity like this:

$$Propensity = Deviation\ (3P\%)^3$$

In the above equation, propensity is defined as a player's deviation times their three-point shooting percentage cubed. By cubing three-point shooting percentage, I ensure that the worst shooters receive a large compounded penalty, while the best shooters receive the smallest penalty. When players are ordered by their propensity, those with the highest scores are both effective and highly utilized, while players with the most negative scores are the least utilized, though still very effective.

# Results

## Make-Miss Model

First, consider the results of the make-miss model. Recall that the make-miss model was trained to measure the relative importance of each feature in predicting a made shot. Figure 5 summarizes the returned Boruta importance scores for each feature relative to each other.

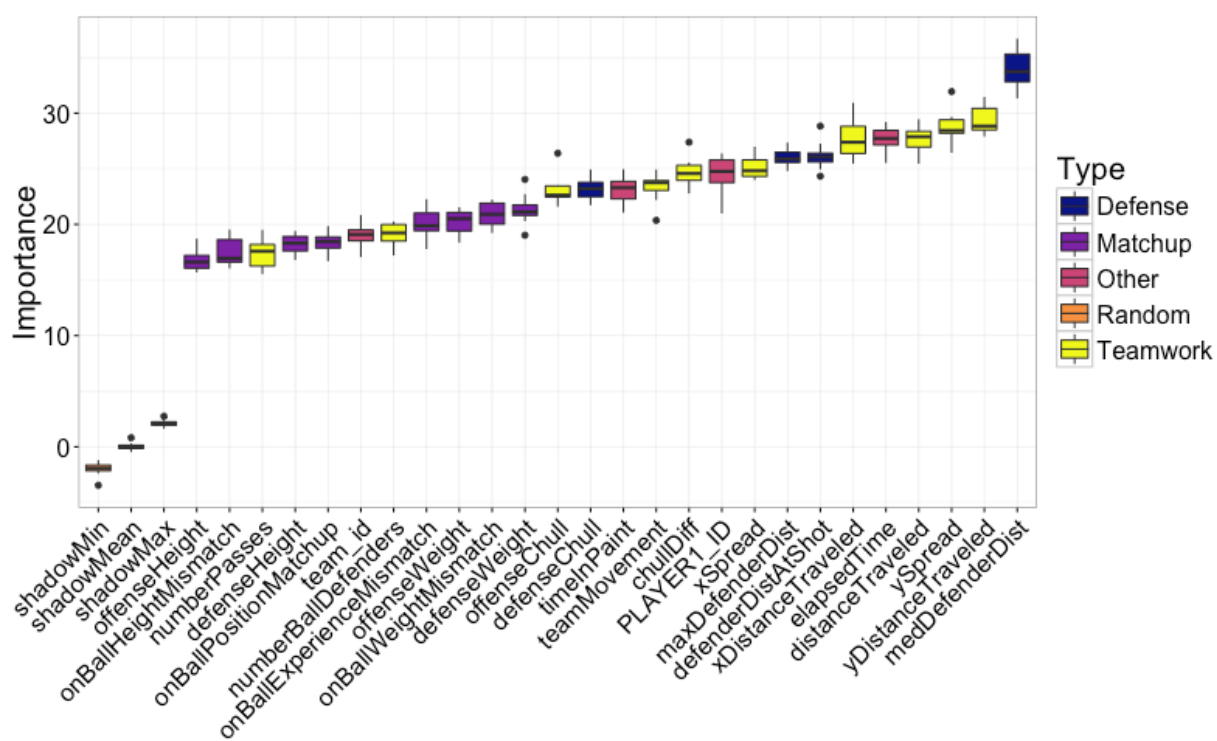

*Figure 5: Boruta feature importance distributions for the make-miss model*

As depicted in Figure 5, the boxes for the shadow variables on the far left-hand side of the figure represent the Boruta importance scores for randomly permuted variables. Because each shadow feature represents the distribution of importance scores for random source feature permutations, we can infer that each of our features is at least more predictive than random.

Other key takeaways from Figure five are that teamwork metrics (e.g., offensive convex hull and ball movement) are generally more predictive of success than player matchups. Unsurprisingly, some of the strongest predictors capture the distance between the shooter and the nearest defender.

The rest of these results can be easy to gloss over, so I will describe some of the more nuanced findings here, and provide context. First, of all the metrics tested, the one that is most predictive of a make or miss is the average (median) distance between the shooter and the closest defender over the course of that play. This result is expected: it is easier for a player to shoot when they are open, and it is more difficult to shoot when they are being defended closely. In a sense, this finding provides a sanity check on the rest of the findings in this analysis.

Next, I want to jump down the list a little to point out PLAYER1_ID. This feature represents the identity of the shooter. Consider what that means in this context. The identity of a player essentially captures the difference in player skill and efficiency in one feature. Its relative importance tells us how significant it is that a good player is shooting vs. a bad one. Furthermore, its position on the list of important features is very noteworthy because there are many features ahead of it. This suggests that many features, such as ball movement, and shot timing, are more predictive of three-point success in the NBA than the shooter's skill.

Next, consider the various features that capture shooter-defender matchups. Many offensive game strategies involve luring the defense into personnel mismatches, through screen setting, or other means. For example, it is generally accepted that big players can over-power smaller defenders on post-up plays. However, it is not as well understood how mismatches can be exploited on three-point shots. According to these results, the difference in height, weight, experience, and position between a shooter and his nearest defender all have relatively low power at predicting three-point shot outcomes when compared to strategies that involve ball movement, court spacing, shot timing.

**Player Model**

Next, consider the results of the player model. Recall the leave-one-out modeling architecture that was deployed for scoring players in the player model, and consider the distribution of model performance observed on each of the player models. Below, I plot a histogram of model performance in terms of $R^2$ and RMSE (root mean squared error) for the test set scored on each of the player models.

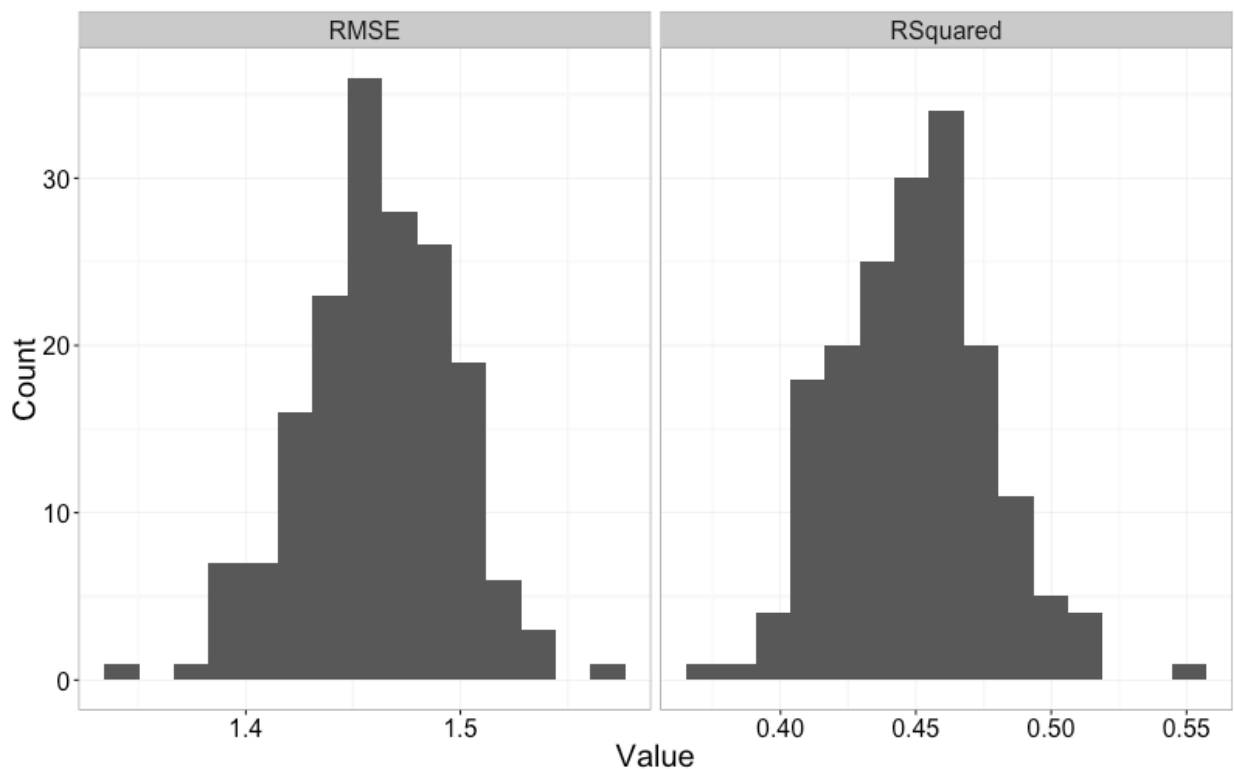

*Figure 6: Histograms of RMSE and R squared across all player models*

In the histograms depicted in Figure 6, we can see that the distribution of player model performance is approximately normal for both RMSE and $R^2$. The narrow shape of each distribution suggests stable model performance across players. Furthermore, we can observe that the models display a reasonable variance; mean $R^2$ is around 0.46, with maximum around 0.55, and minimum around 0.39. These results should offer confidence in the stability of performance across player models.

Next, recall that the player model aggregates the features derived in the make-miss model to each player in the dataset for a comprehensive measurement of player and team movement during three-point shots. The player model uses these aggregate features to infer each player's per-game three-point attempt rate. By comparing a player's model-inferred three-point attempt rate to their actual three-point attempt rate, we can observe players who behave in unique ways in terms of their three-point shooting. Specifically, this comparison allows us to deduce if a given player shoots more frequently or less frequently than would be expected of another player in their situation. Consider first players who shot more threes per game than expected:

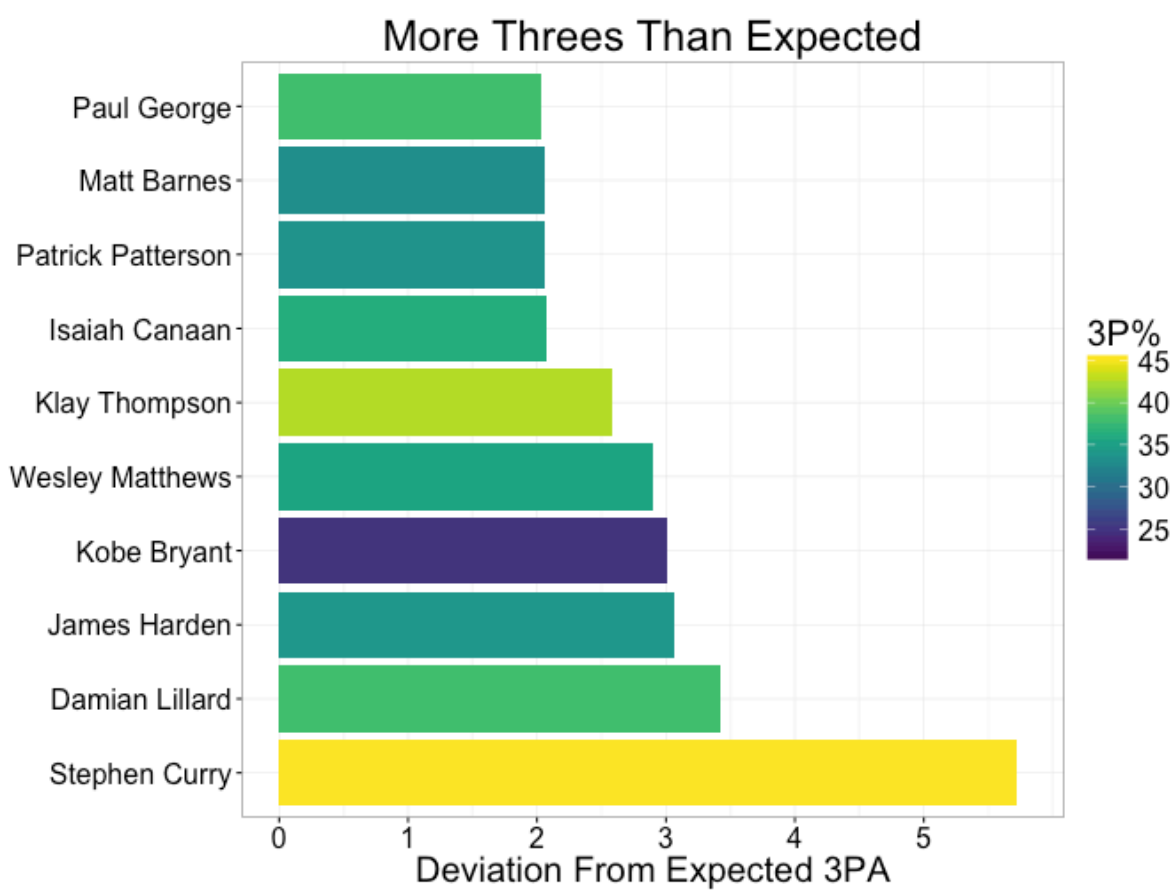

*Figure 7: Players who shot more threes than their model expected, colored by their respective three-point shooting percentage*

In Figure 7, the size of the bar associated with each player corresponds to the deviation of their actual three-point attempt rate from their model-expected three-point attempt rate (more three-point attempts than expected).  The color of each bar offers context by conveying the three-point shooting percentage of the corresponding player.  The players shown in Figure 7 are the top ten positive deviators from their model's projection. We can see that Stephen Curry averaged 5.7 more three-point attempts per game than expected and is also a very efficient three-point shooter.  Given the high efficiency of the two-time most valuable player, he should be a welcome outlier.

Conversely, we see that Kobe Bryant averaged 3 more three-point attempts per game than his model expected, but was a very inefficient three-point shooter.  Knowing his specific situation is revealing; 2015-16 was the final season of Bryant's long and storied career.  Though these results suggest he was forcing up many more three-pointers than other players in his position would have, his team presumably put up with such inefficient performance in honor of his final professional season, and to give their fans a final glimpse of him in action.  Next, consider players who shot fewer threes per game than expected.

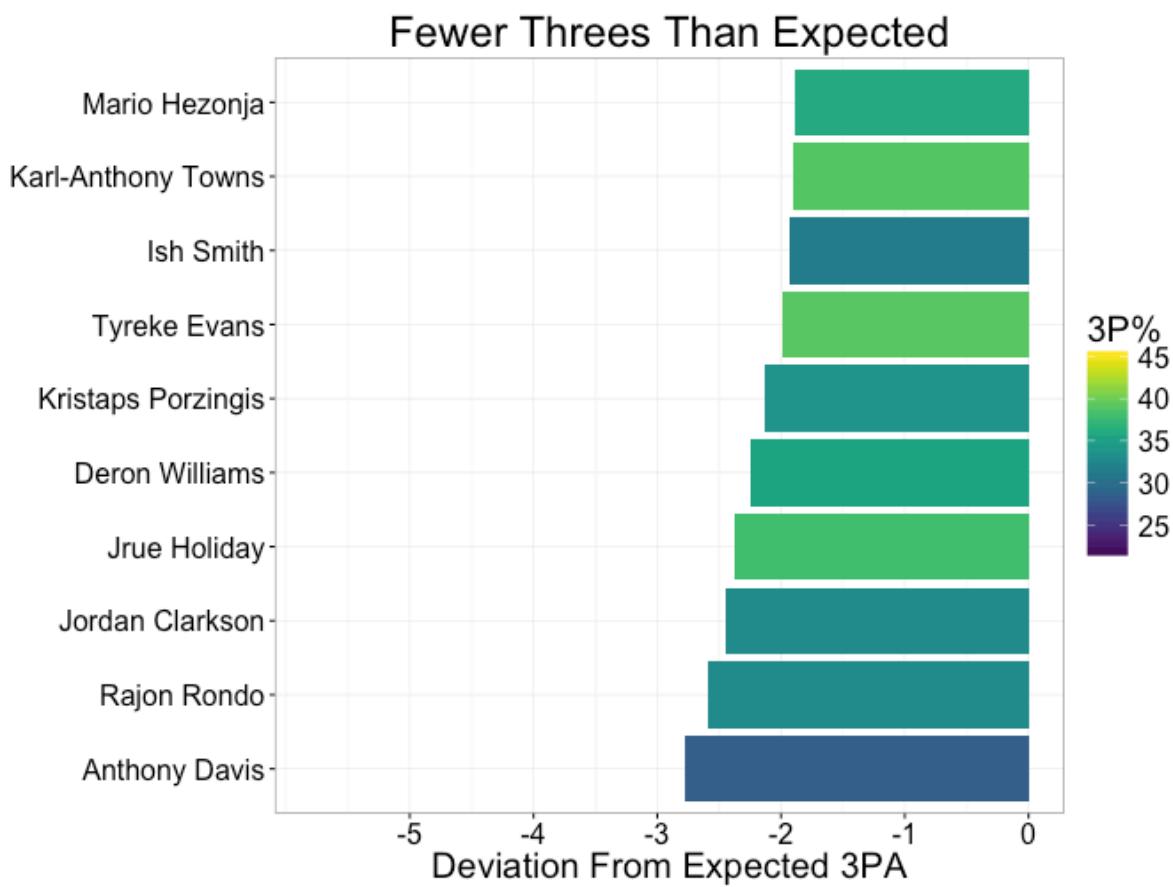

*Figure 8: Players who shot fewer threes than their model expected, colored by their respective three-point shooting percentage*

In Figure 8, we see the top ten negative deviators from their model's projection.  We can observe that Karl-Anthony Towns averages 2 fewer three-point attempts than expected.   Given the relatively high efficiency with which he shoots the three from the center position, it would be a promising strategy to stretch him out to the three-point line more often.  Conversely, though my model projects Anthony Davis to shoot 2.8 more threes per game than he really did, his mediocre shooting percentage does not warrant a game strategy where he takes too many more three-point shots.

The results illustrated in Figures 7 and 8 and discussed above are very telling. They highlight effective and ineffective shooters in the context of how other players would perform in their situation. However, they do not convey the whole story. As demonstrated in this discussion, there is a meaningful relationship between a player's deviation from model-expected three-point attempts and their three-point shooting percentage. Likewise, I defined the propensity metric for measuring this relationship. Recall that when players are ordered by their propensity, those with the highest scores are both effective and highly utilized; these players consistently make shots that their peers would not. Conversely, players with the most negative propensity scores are very effective shooters who are under-utilized; they represent players with the most missed opportunities; despite being effective shooters, they refrain from shooting more often than their peers would in similar situations. Consider the players with the strongest propensity from each of these two groups (effective high-utilization and effective low-utilization), as listed in Figure 9.

**Effective, High-utilization**

| Player | Propensity |
|--------|------------|
| Stephen Curry | 0.5250 |
| Klay Thompson | 0.1984 |
| Damian Lillard | 0.1881 |
| Wesley Matthews | 0.1310 |
| James Harden | 0.1206 |
| Hollis Thompson | 0.1156 |
| Paul George | 0.1119 |
| Kyle Lowry | 0.1075 |
| J.R. Smith | 0.1008 |
| Isaiah Canaan | 0.0989 |

...

| | |
|--------|------------|
| Jeff Teague | -0.0939 |
| Troy Daniels | -0.0974 |
| Chris Paul | -0.0977 |
| Deron Williams | -0.1005 |
| Ian Clark | -0.1005 |
| Kawhi Leonard | -0.1023 |
| Karl-Anthony Towns | -0.1117 |
| Tyreke Evans | -0.1179 |
| Jrue Holiday | -0.1290 |
| Luis Scola | -0.1468 |

**Effective, Low Utilization**

*Figure 9: Three-point shooters, ordered and colored by their three-point shooting propensity*

The most effective, highly utilized players are observed in the first table of Figure 9. The list includes many household names, such as the historically great shooter and MVP Stephen Curry, his teammate Klay Thompsan, as well as Damian Lillard, James Harden, and Paul George. These players' label as great shooters will be no surprise to NBA fans. However, when assessed by the same standards, several other lesser-heralded shooters rank highly; Wesley Matthews, Hollis Thompsan, and Isaiah Canaan are all well regarded shooters, but rarely have their three-point shooting prowess compared to the superstars cited above.

Similarly, the second table of Figure 9 lists the most effective and under-utilized three-point shooters. Again, this list is of particular interest because it calls to light players who could expect to be successful if they shoot more three-pointers. As before, we see the rookie-of-the-year center Karl-Anthony Towns with a strong ranking by this metric. In short, this is a significant list because these players have unlocked potential in terms of three-point shooting. Knowing this, teams can adjust game strategy around these players, or target under-the-radar players for sneaky talent acquisition.

# Conclusions

In my thesis, I measure the relative influence of competing basketball game strategies, and I apply my analysis to plays resulting in three-point shots. I use SportVU player tracking data from NBA games to derive custom position-based features that capture tangible game strategies from game-play data. Then, I demonstrate statistical methods for measuring the relative importance of any given basketball strategy. In doing so, I highlight the high importance of teamwork based strategies in affecting three-point shot success. By coupling the most recent innovation in basketball data-gathering (SportVU), with an advanced variable importance algorithm (Boruta), I am able to extract meaningful results that were not feasible even 3 years ago. Furthermore, I demonstrate how player-tracking based features can be used to measure the three-point shooting propensity of players, and I show how this measurement can identify effective shooters that are either highly-utilized or under-utilized. Altogether, these findings provide a substantial body of work for influencing basketball strategy, and for measuring the quality of basketball players.

Though three-point shooting was the focus of my research, that choice was an arbitrary one to narrow my scope. The methods I demonstrate in my research can be applied to a number of game targets as long as they can be measured (i.e. 2-point shooting, pick-and rolls, team rebounding, defense, etc.). Similarly, the features that I define in the make-miss model were also only arbitrary selections based on quantifiable game strategies; any game strategy can be tested in this framework as long as it can be measured.

In the player model, I construct a highly meaningful model that was trained only on the features defined for the make-miss model. However, these features are limited in their ability to capture relevant game-play information, and their explicit definitions are not relevant for the player model's utilization. Likewise, a more encompassing approach to training a player model would be based on a neural network style architecture. The benefit of a neural network in this situation is that it can take the raw player tracking data as inputs, and automatically learn the relevant features and interactions for a given target (i.e. three-point shooting). One could thus expect a neural network style model to achieve even better performance than the model I demonstrate in this research. Moreover, as discussed in the literature review, neural networks have already been successfully demonstrated for use-cases on the NBA player tracking data.

In close, my work pushes the envelope for analyzing basketball strategy, and for measuring the quality of basketball players. Until recently, the analyses demonstrated in this paper were not even feasible. They were only made possible with the availability of player tracking data and with the latest advances in statistical learning. Much is still yet to be done to advance both my work and the field of basketball analytics as a whole. SportVU data has opened many new doors for basketball analytics, and each new analysis snowballs many more questions about our perception of the game.